\documentclass[a4paper,conference]{IEEEtran}
\ifCLASSINFOpdf
  \usepackage[pdftex]{graphicx}
\else
  \usepackage[dvips]{graphicx}
\fi
\usepackage{caption}
\usepackage{subcaption}
\usepackage{amsmath}

\begin{document}
%
\title{No-Sense: Sense with Dormant Sensors}

\author{\IEEEauthorblockN{Chayan Sarkar, Vijay S. Rao, R. Venkatesha Prasad}
\IEEEauthorblockA{Embedded Software Group\\
Delft University of Technology, The Netherlands\\
\{C.Sarkar, V.Rao, R.R.VenkateshaPrasad\}@tudelft.nl
}}


%



\maketitle

\begin{abstract}
The lifetime of a wireless sensor network mainly depends on battery capacity and energy consumption at each node for operations such as, sensing, processing and communication. Popular approaches to save energy have been to intelligently duty cycle and restrict the frequency of these operations, rendering lower quality data at the sink. In this article, we propose \emph{Virtual Sensing Framework} (VSF), which reduces the frequency of the above mentioned operations at each node while not compromising on the sensing interval, and hence resulting in higher quality data at the sink. VSF creates virtual sensors at the sink to exploit the spatio-temporal correlations among sensed data. Using an adaptive model at every sensing iteration, the virtual sensors can predict \textit{multiple consecutive} sensor data for dormant physical sensors with the help of only a few active physical sensors. We show that even when the sensed data represents different parameters (e.g., light, temperature), our proposed technique works well. Applying our technique on the real-world data sets, we attain substantial reduction in energy consumption per node while maintaining high accuracy of the sensed data. To achieve higher energy reduction, VSF has to be used in conjunction with various layers and protocols of the communication stack. Thus, it has the potential to open up new research insights to make the best use of statistical properties of collected sensor data in a network.
\end{abstract}


%
\IEEEpeerreviewmaketitle

\section{Introduction}
\label{intro}
Wireless sensor networks (WSNs) have enabled continuous monitoring of an area of interest (body, room, region, etc.) while eliminating expensive wired infrastructure. Typically in such applications, wireless sensor nodes report the sensed values to a sink node, where the information is required for the end-user. WSNs also provide the flexibility to the end-user for choosing several parameters for the monitoring application. For example, placement of sensors, frequency of sensing and transmission of those sensed data. Over the years, the advancement in embedded technology has led to increased processing power and memory capacity of these battery powered devices. However, batteries can only supply limited energy, thus limiting the lifetime of the network. In order to prolong the lifetime of the deployment, various efforts have been made to improve the battery technologies and also reduce the energy consumption of the sensor node at various layers in the networking stack. Of all the operations in the network stack, wireless data transmission and reception have found to consume most of the energy. Hence many proposals found in the literature target reducing them through intelligent schemes like power control, reducing retransmissions, etc. In this article we propose a new framework called {\it Virtual Sensing Framework} (VSF), which aims to sufficiently satisfy application requirements while conserving energy at the sensor nodes.
\begin{figure}
\centering
	\begin{subfigure}[b]{0.42\linewidth}
		\includegraphics[width=\textwidth]{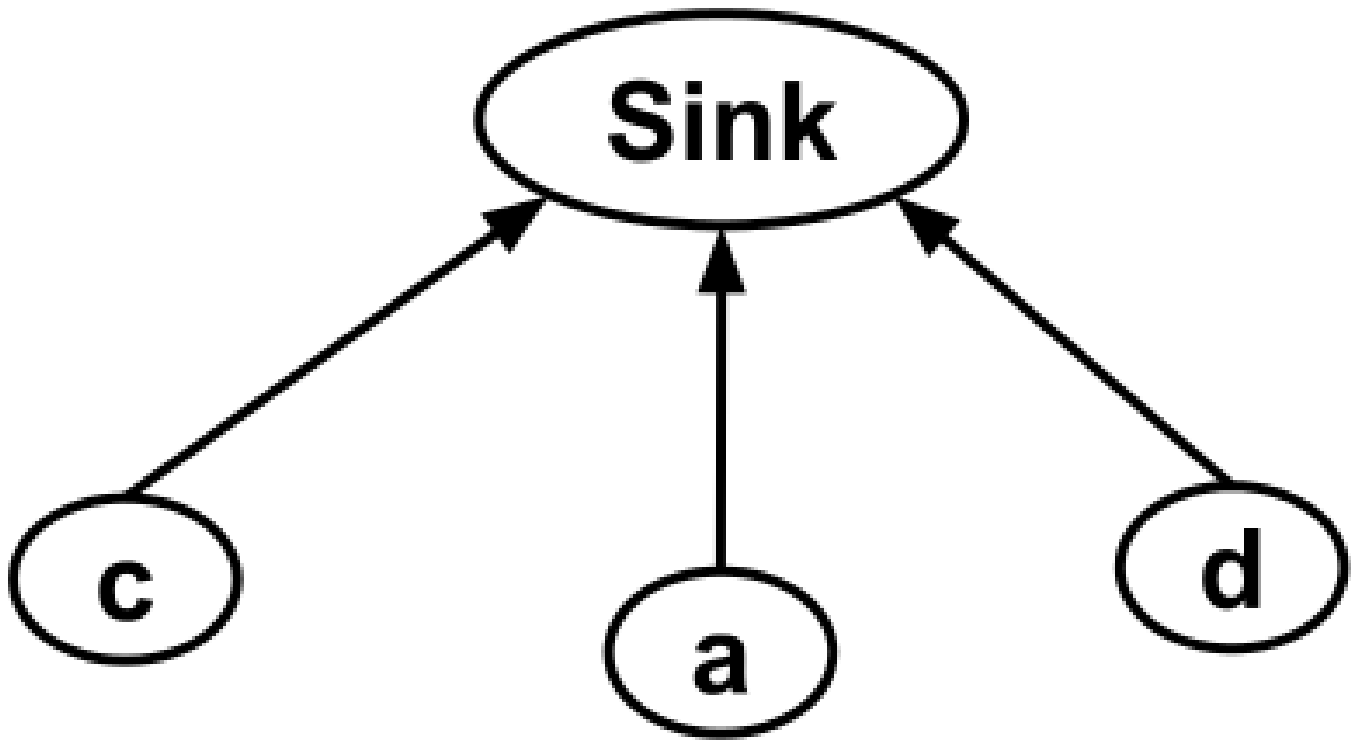}
		\caption{}
		\label{fig:scenario1}
	\end{subfigure}
\hspace{2em}
	\begin{subfigure}[b]{0.42\linewidth}
		\includegraphics[width=\textwidth]{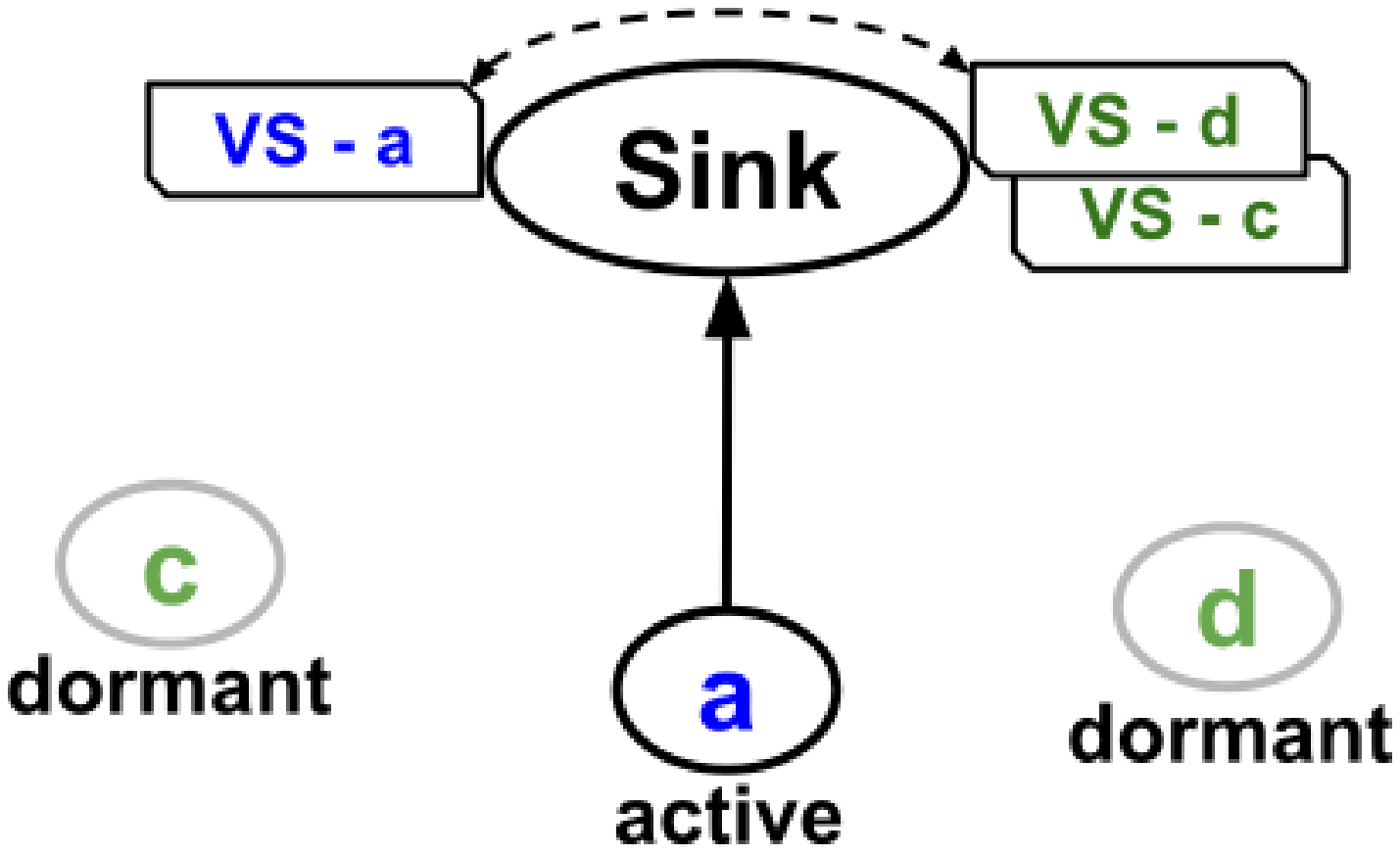}
		\caption{}
		\label{fig:scenario2}
	\end{subfigure}
\caption{(a) Data collection scenario in a WSN (star topology); (b) Data collection with virtual sensing framework.} 
\label{fig:wsn}
\end{figure}

In VSF, we define a {\it Virtual Sensor} (VS) as a software construct in the sink that represents a Physical Sensor (PS)\footnote{A physical sensor is nothing but an wireless sensor node in the WSN.} in the WSN. A PS can either be dormant (lowest power mode or sleep mode) or be active (microcontroller, sensors and/or radio are on). When a PS is active, the frequency of its sensor data reporting varies adaptively, which is described later. A VS holds the measured sensor data reported by its corresponding PS, when active, or predicts sensor data on behalf of its PS, when dormant, by exploiting the temporal and spatial correlations among {sensors\textquoteright} data in the WSN. It is important to note that the VSF does not consider any \textit{a priori} knowledge about the correlation patterns in the data collected from the sensors. Rather, it utilizes adaptive prediction schemes for each sensor based on inherent correlation among the {sensors\textquoteright} data.  This is known as non-model approach \cite{santini2006adaptive}. 

Each VS saves energy for its PS by following a collaborative technique to balance between the accuracy of the predicted values and the energy consumption at its PS. Since VSs {\textquotedblleft reside\textquotedblright} in the sink/base station, the collaboration has insignificant overheads; the only overhead is notifying the PSs whether to be in sleep mode, or in active mode for a certain duration. 

The concept of VSF is depicted in Fig.~\ref{fig:wsn}. Here, we have considered one hop to describe VSF and its mechanisms, and illustrate the working of VSFs through real-data. It is possible to extend the work to a multihop scenario, which we consider as future work. Nevertheless, several WSNs applications can make use of VSF - many periodically reporting WSN can utilize VSF to conserve energy. A case in point is smart buildings, where sensors monitor humidity, ventilation and air conditioning (HVAC) continuously. With VSF, some sensors need not transmit data until their values have changed significantly. With this background we list below the major contributions of this work: 
\begin{itemize}
\item To the best of our knowledge, we are the first to propose a generic energy saving scheme for sensors based on data prediction that can be applied to any type of data collection network. The generality of the method lies in prediction of \emph{multiple consecutive} sensor values for dormant sensors without having prior knowledge (or model) of the sensed data, physical nature of the sensing and deployment information of the nodes. 
\item Our proposal exploits both the spatial and temporal correlations among sensors. It can track the changes in the correlation and adapts to the situation through the use of blind adaptive filtering technique. 
\item The energy saving technique proposed in this article ensures that every node spends energy almost evenly. We show that the overall energy consumption of the whole deployment is reduced, while maintaining sufficiently high accuracy of the predicted sensed values. 
\item We propose to predict consecutive sensor data to keep the sensor nodes dormant for a longer duration whereby reducing the overhead of switching between low power and active modes.
\end{itemize}
%
\section{Related work}
\label{sec:rel}
While many energy saving techniques/protocols have been proposed for wireless sensor networks, we only review the works that are relevant to our proposal. One common approach to reduce energy consumption is to select a subset of nodes among all the sensors deployed in the network. As the sensors usually have spatial correlation, missing sensor data could be reconstructed from the subset of sensor data. Most of the proposals found in the literature do not find the correlation among sensors explicitly. Rather a good correlation structure among the sensors is assumed to be known \textit{a priori} \cite{chu2006approximate,cristescu2005optimal, deshpande2004model, Gupta2008, jain2004adaptive, Quan2007}. However, we do not assume any predefined correlation structure and our adaptive scheme tracks the correlation amongst the sensors in real-time. Further, it should be noted that, to support our prediction, we do not draw any inference based on the physical nature of sensing and deployment of sensors. 

A LMS-based adaptive prediction technique has been exploited by Santini {\it et al.} \cite{santini2006adaptive}, which does not consider prior knowledge about the sensor data. While our strategy is similar, we exploit spatial correlation to estimate the sensing values blindly. That is, our scheme can predict multiple consecutive sensing values while the sensor node remains dormant. A real-time, blind prediction scheme has also been proposed by Li {\it et al.} \cite{Li2008}. However, they assume predefined spatial correlations among the sensors and it is required to have a very high spatial correlation between the sensors. 

Guestrin {\it et al.} \cite{Guestrin2004} provided a technique where sensor nodes save energy by restricting their data transmission. The nodes only transmit the parameters of estimation model -- instead of the data itself -- only if significant changes are noticed in the sensor data. In contrast with this method, our approach can predict consecutive sensor values accurately while the node is in sleep mode. There are some works where estimation techniques have been used to predict sensor data in order to fill the missing data points and complete the sensor data set \cite{Pan2010}. Thus, they are not suitable for online sensor data prediction for consecutive sensing instances.

\section{Virtual Sensing Framework (VSF)}
\label{sec:vsp}
VSF aims to reduce energy consumption of a sensor network by reducing its activity, i.e., reducing frequency of sensing, processing and data transmissions. The data collection is complemented by predicting sensor data at the sink. To predict the sensor data, a virtual sensor (VS) is created for each physical sensor (PS) in a deployment at the sink as shown in Fig.~\ref{fig:wsn}. A VS instructs its associated PS on its state and activity for the forthcoming sensing intervals. VSs also collaborate among themselves in order to save more energy and maintain high data accuracy. In the subsequent sections, we discuss how sensors collaborate among themselves and then discuss how the prediction is done.

\subsection{Energy saving technique}
\label{policy}
Usually physical parameter values (e.g., ambient temperature) do not change abruptly in a short time span.  Therefore, these values have correlation with their immediate past values (temporal correlation). Thus, a sensor value can be predicted by exploiting its temporally correlated data. In order to increase the energy savings, VS should predict successive values while its PS remains dormant. As a result, changes in the physical parameter, during the longer dormant periods, might not be captured by the temporal correlation based method. Prediction in this case can be improved with spatial correlations: if two sensors have had very high correlation in the recent past, it is safe to say that both the sensors will behave in a similar fashion for some time in the future too. Hence in VSF, we choose to exploit temporal as well as spatial correlations in the data collected from the sensors to fine-tune the prediction. With increased accuracy of predictions and correlated sensors, one sensor can remain dormant for the duration of prediction with the help of the other sensor. The latter sensor \textit{aka} the helper is referred to as {\it companion}. The companion, therefore, has to be an active PS. Please note that the companion of a dormant sensor is not predefined and fixated i.e., it can change over time based on changing correlation between the sensors.

VSF also conserves energy in the companion PS, whenever possible, by incorporating a temporal correlation based predictor within it, and VS in the sink. Here the node continues to sense the physical parameter, and also predicts the value. If the prediction error lies within a sufficiently tolerable error bounds, then the sensor does not transmit the sensed data. By withholding data transmissions, significant energy is saved even in the active sensors. The dormant and active nodes are associated with two different types of VSs based on their functionality. We use the terms Type-I VS when the corresponding PS is in dormant state,  and Type-II VS when the corresponding PS is active. Note that Type-II VS may or may not send the sensed data depending on the prediction error sought. 

Every Type-I VS requires at least one companion, while one companion can help multiple Type-I VS. Multiple spatially correlated sensors can act as companion nodes for a Type-I VS to improve the prediction accuracy, however after a certain threshold number of companions, more such companions will not necessarily improve the prediction significantly. This trade-off between prediction accuracy and energy savings (more active node implies more energy consumption) needs to be balanced. We follow a greedy approach to maximize the energy saving: VSF maximizes the number of Type-I VS nodes, and thus reducing the number of companions.

It is clear that a dormant sensor can conserve more energy than an active sensor. In order to keep the sensor network alive for a longer period, energy consumption of the sensor nodes in the network need to be balanced over time.  Thus, state of the nodes switch between dormant and active after certain number of timeslots. 

\begin{figure}
\centering
\includegraphics[width=\linewidth]{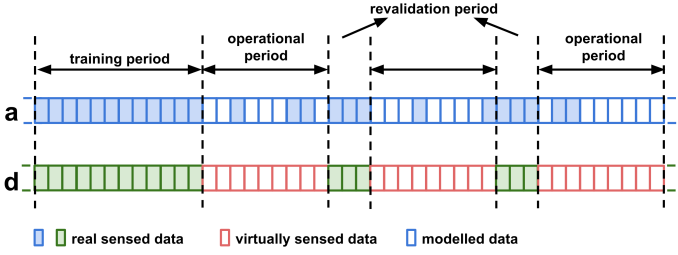}
\caption{Data collection phases in virtual sensing technique.}
\label{fig:vs_train}
\end{figure}

As we do not assume any \textit{a priori} knowledge about the sensor data statistics, VSF needs to capture the correlation among sensors. It should also monitor the change in the correlation and adapt dynamically. To accomplish this, the whole data collection period is classified into three phases -- training period, operational period and revalidation period as shown in Fig~\ref{fig:vs_train}. During the training period, all the PSs collect data and transmit their data to the sink. Using the training data sets, states of the nodes (active or dormant), suitable companion for each dormant node, type of their VSs, and prediction models for each VS are decided. During operational period, the VSs become functional as the dormant and active nodes restrict their activity to conserve energy. The revalidation period resumes all PSs to active mode for a shorter period of time as compared to the training period. The prediction models are validated using the sensor data collected in this period, and the operational period resumes. If the revalidation (and calibration) of the prediction model fails, i.e., the correlations among the sensor data changes drastically, another training period is initiated. The following sections describe each of the phases in detail.

\subsection{Training period}
During the training period, $T_{p}$ training data samples ($T_{p} > 0$) are collected from all the sensors. Using these samples, a temporal prediction mechanism is created for each type of VS at the end of the training period. To create a prediction mechanism based on temporal correlation, a transversal or tapped-delay line filter is created. First, the training data is stored in a $(T_{p}-p)\times p$ matrix and $(T_{p}-p)\times 1$ column vector as inputs ($ U $) and outputs ($\underline{d}$) of the filter as shown below, 
\begin{alignat}{2}
U & = \left[ \begin{array}{cccc}
d(p) & d(p-1) & \cdots & d(1) \\
d(p+1) & d(p) & \cdots & d(2) \\
\vdots & \vdots & \ddots & \vdots \\
d(T_{p}-1) & d(T_{p}-2) & \cdots & d(T_{p}-p)
\end{array} \right] \\
\underline{d} & = [d(p+1), d(p+2), ..., d(T_{p})]^{T}, 
\end{alignat}
where $p$ is the order of the filter. Then, the filter coefficients, 
\begin{equation}
\underline{\alpha} = (U^{T}U)^{-1}U^{T}\underline{d}, \label{eq:filcoeff}
\end{equation}
are found by minimizing the mean-square error~\cite{haykin2005adaptive}. These $p$ filter coefficients, i.e., $\underline{\alpha} = [\alpha_{1}, \alpha_{2}, ..., \alpha_{p}]^{T}$ are used to predict the sensor data during the operational period. If the correlation is known a priori, a {\it Wiener filter} can be developed, which is said to be the {\it optimum in the mean-square error sense}. As the autocorrelation function is unknown, the filter coefficients can become outdated and may result in erroneous prediction. Using an adaptive filtering technique, the coefficients are updated at later stages.

A Type-I VS also needs to develop a prediction mechanism based on spatial correlation. To calculate a spatial correlation based predictor, VSF uses linear regression analysis. Linear regression is a statistical method that models the relationship between a dependent variable and one/more independent variable(s)~\cite{montgomery2012introduction}. In our approach, we treat the dormant node as dependent variable and a potential companion as an independent variable. The inputs for linear regressor are collected from the companion node and stored in the form,
\begin{equation}
V  =  \left[ \begin{array}{cc}
1 & a(p+1) \\
1 & a(p+2) \\
\vdots & \vdots \\
1 & a(T_{p})
\end{array} \right] .
\end{equation}
Then, by minimizing the mean-square error, we calculate the coefficient of the linear regression as,
\begin{equation}
\underline{\beta}  =  (V^{T}V)^{-1}V^{T}\underline{d}. \label{eq:beta_calcu}
\end{equation}
The vector $\underline{d}$ is used as the regressor output. $\underline{\beta} = [\beta_{0}, \beta_{1}]^{T}$ are used for spatial prediction during the operational period, and they are also updated during the revalidation period (discussed in Section~\ref{sec:rvp}). 
\subsubsection*{Finding a suitable companion}
Since we do not assume any prior spatial correlation, a companion cannot be chosen beforehand. A common assumption is that two geographically co-located sensors show high spatial correlation. Nevertheless, in reality, they may show poor correlation sometimes, whereas two sensors located relatively far can show high correlation. By finding a suitable companion, we find a spatially correlated sensor that can best predict the sensor data. Again, since we do not consider any prior knowledge about the physical parameters of the sensed data, we can indeed make any sensor a companion for any other sensor during the training period. The linear regression coefficients are created separately for each potential companion using (\ref{eq:beta_calcu}). The sensor, which scores the highest in the {\it goodness of fit} test, is selected as the companion.
{\it Chi-squared statistics} is a well-known method to test goodness of fit~\cite{taylor1997introduction}. To this end, the error values of the estimated signal need to be known. Using the model parameters and the training data set, first, the sensor values are estimated as ${\underline{d}}_{spa} = V\underline{\beta}$. Then, the Chi-squared statistics can be obtained by taking normalized sum of the squared-errors. Chi-squared statistic is calculated as,
\begin{equation}
\chi_{spa}^{2} = \displaystyle\sum_{i=p+1}^{T_{p}}\frac{(d(i) - {d}_{spa}(i))^{2}}{\sigma^{2}}, 
\label{eq:spa}
\end{equation}
where $\sigma^{2}$ is the variance of the observed signal. To get an inference from the statistics, a reduced Chi-squared statistic can be calculated by dividing it by the number of degrees of freedom. The score of the goodness of fit test, represented as $\delta$, is given by, 
\begin{equation}
\delta = 1 - \frac{\chi_{spa}^{2}}{\nu},
\label{eq:del}
\end{equation}
where, $\nu$, the degrees of freedom is equivalent to the number of samples ($T_{p}-1$). $\delta$ lies between $(0, 1)$, where 0 implies complete failure of capturing the system behavior and 1 implies complete resemblance of the system behavior by the model parameters.
\begin{figure}[]
\centering
\includegraphics[scale=0.15]{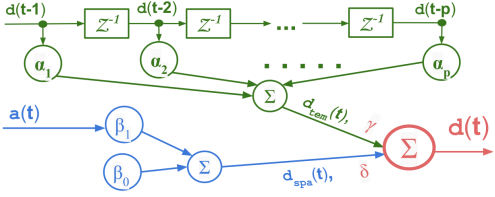}
\caption{The filter used for an active virtual sensor is a hybrid of a transversal filter (for temporal prediction) and linear regression (spatial prediction).}
\label{fig:filter}
\end{figure}
\subsection{Operational period}
Once the prediction coefficients are found, the operational period starts (see Fig.~\ref{fig:vs_train}). In this period, a Type-I VS informs its corresponding PS to remain dormant for $O_{p}$ timeslots, and it predicts consecutive $O_{p}$ data points using a hybrid model consisting of temporal and spatial predictions (see Fig.~\ref{fig:filter}).
At $t^{th}$ timeslot, first, the temporal prediction is done using
\begin{equation}
d_{tem}(t) = \displaystyle\sum_{i=1}^{p}\alpha_{i}\cdot d(t-i),
\end{equation}
where $\alpha_{i}$ is found using (\ref{eq:filcoeff}). Then the VS collects the sensor data from its companion. Using the data from the companion and the linear regression coefficients, the spatial prediction is found using 
\begin{equation}
d_{spa}(t)  = \beta_{0} + \beta_{1}\cdot a(t).
\end{equation}
The final predicted value of the missing sensor data is computed by taking a weighted average of these two predicted values as,
\begin{equation}
\hat{d}(t)  =  \frac{(\gamma\cdot d_{tem}(t) + \delta\cdot d_{spa}(t))}{(\gamma + \delta)}. 
\label{eqn:gamdel}
\end{equation}
The weights $\gamma$ and $\delta$ are the {\it goodness of fit} score of the temporal and spatial predictions respectively. The value of $ \gamma $ is calculated the same way as $ \delta $ has been calculated while choosing the best companion using (\ref{eq:spa}) and (\ref{eq:del}).

During the operational period, a Type-II VS informs its corresponding PS to continue its sensing measurement for the predefined interval and to avoid any unnecessary data transmissions if possible. At the beginning of the operational period, Type-II VS sends the filter coefficients to the physical sensor. At $t^{th}$ instant in operational period, the sensor predicts the sensor value using the filter coefficients, $\underline{\alpha}$, (see (\ref{eq:filcoeff})). Moreover, the actual measurement is also available since it is active. Then, it calculates the prediction error as given by (\ref{eq:error}) using the model parameters. If the absolute error of the prediction is within a permissible (predefined) limit, then sensor node suspends transmission and the corresponding Type-II VS predicts the sensed value using the same filter coefficients. Otherwise, the sensor node transmits the value to the sink and the model coefficients are updated both at the sensor node and at the corresponding VS in the sink. In this way, the model parameters are synchronized at both the places. At the same time, we can ensure that the prediction error remains within the permissible error limits. To update the model parameters (at both the places), we have used {\it least-mean-square (LMS) algorithm}. LMS is a widely used adaptive filtering technique in time-series predictions. It requires very less memory and computational capabilities, and performs well~\cite{santini2006adaptive}.
 
To update the temporal filter coefficients, i.e., the $\underline{\alpha}$ vector, first, the prediction error is calculated given by,
\begin{equation}
e_{temp}(t)  =  d(t) - d_{temp}(t), \label{eq:error}
\end{equation}
where $d(t)$ and $d_{temp}(t)$ are the actual and predicted sensor values respectively. Then the filter coefficients are updated using,
\begin{equation}
\underline{\alpha}(t+1)  =  \underline{\alpha}(t) + \mu \cdot \underline{d}(t) \cdot e_{temp}(t), \label{eq:model_updat}
\end{equation}
where $\underline{d}(t) = [d(t-1), d(t-2), ..., d(t-p)]^{T}$ is the input vector of the filter, and $\mu$ is the learning rate of the adaptive algorithm~\cite{haykin2005adaptive}. Procedure to set $\mu$ can be found in~\cite{santini2006adaptive}. 
\subsection{Revalidation period}
\label{sec:rvp}
Since Type-I VS asks the corresponding PS to go to dormant mode, there is a chance that the predicted value might diverge from the ground reality. To tackle this, revalidation of the model parameters are done after $O_{p}$ sensing intervals (here we skip the discussion on fixing $O_{p}$ due to paucity of space). During this period, i.e., $R_{p} (<T_{p})$ sensing intervals, all the PSs becomes active. That is, they sense and transmit the data to the sink (see Fig.~\ref{fig:vs_train}). Then, the temporal prediction coefficients are updated as described in (\ref{eq:error}) \& (\ref{eq:model_updat}). The spatial prediction coefficient, i.e., $\underline{\beta} = [\beta_{0}, \beta_{1}]^{T}$, is also updated based on the spatial prediction error.
As the final prediction of a Type-I VS is dependent on $\gamma$ and $\delta$ (\ref{eqn:gamdel}), an updated {\it goodness of fit} score, i.e., the {\it Chi-squared statistics} is calculated using,
\begin{eqnarray}
\chi_{tem}^{2} & = & \chi_{tem}^{2} + \frac{e_{tem}^{2}(t)}{\sigma^{2}}  \\
\chi_{spa}^{2} & = & \chi_{spa}^{2} + \frac{e_{spa}^{2}(t)}{\sigma^{2}}. 
\end{eqnarray}

\section{Evaluation}
\label{eva}
In this section, we evaluate our virtual sensing technique on the real data sets obtained from the Lausanne Urban Canopy Experiment~\cite{sensorscope}. The data were collected in the EPFL campus between July 2006 and May 2007, where 97 sensor nodes monitored various environmental parameters, e.g., ambient temperature, solar radiation, relative Humidity,, etc. These sensor nodes collected data every 30\,s. We have applied VSF on the ambient temperature data collected from multiple sensor nodes. We provide the most interesting results here. We chose two nodes randomly (which turned out to be node 3 and node 44), and we associated Type-I and Type-II virtual sensors with them respectively. We tested our prediction technique in Matlab.
\begin{figure}[]
\centering
\includegraphics[width=\linewidth,height=2.1in]{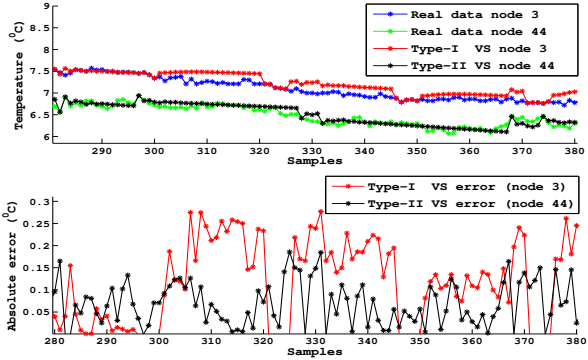}
\caption{Absolute error in sensor data prediction using a Type-I and Type-II virtual sensor.}
\label{fig:prediction}
\end{figure}
\begin{figure}[]
\centering
\includegraphics[width=\linewidth]{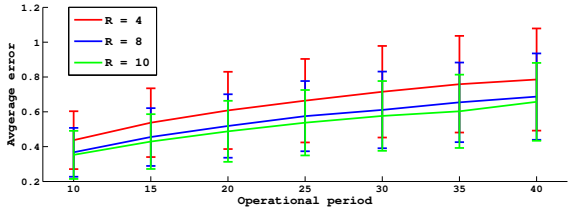}
\caption{Average prediction error for various operational period (M) and revalidation period (R).}
\label{fig:errorplot}
\end{figure}
$10^3$ data samples from nodes are used. A snapshot of the predicted sensor data by the VSs are shown in Fig.~\ref{fig:prediction}. We fixed the operational ($O_{p}$) and revalidation ($R_{p}$) window sizes to 20 and 5 respectively.  The training was done at the beginning of the data streams, which is not shown in the figure. The result shows that the virtual sensors can achieve significant prediction accuracy.

To further investigate the accuracy of the VSF based predictions, we have conducted simulations by varying $O_{p}$ and $R_{p}$. The average error with the standard deviation of error as a confidence interval for various values of the $<O_{p},R_{p}>$ pair is shown in Fig.~\ref{fig:errorplot}. From this study, as expected, when $R$ increases, the average prediction error as well as the variance of prediction error is reduced. Nonetheless, it is at the cost of energy consumption since more data is transmitted. To tackle this we can also increase $O_{p}$. However, with increased $O_{p}$, the average prediction error also increases. The choice of these window sizes depends on the applications as accuracy and energy cost go together. From our simulation results, we can conclude that VSF provides the tool to restrict data transmissions from sensor nodes with an insignificant reduction in the quality of data.

Next, we show that the virtual sensing technique can save more energy using blind estimation. Three simulations are conducted for three different lengths of $O_{p}$ and $R_{p}$. For each pair of $O_{p}$ and $R_{p}$, the error threshold is also varied from $0^{\circ}$C to $2^{\circ}$C. Error threshold 0 represents the base case, where all 1000 data samples from the sensors are sensed and transmitted. 
\begin{table}
\centering
\caption{System Parameters and Settings}
\begin{tabular}{|l|l|} \hline
\textbf{Parameter} & \textbf{Value} \\ \hline
Message size & 128\,B \\ \hline
Transmission power & 0\,dBm \\ \hline
Energy cost for sending a message & 341\,$\mu$J \\ \hline
Energy cost for sensing temperature & 330\,$\mu$J \\ \hline
Energy cost in active mode & 4.898\,mW \\ \hline
Energy cost in low power mode (LPM3) & 0.144\,mW \\ \hline
Energy cost in switching (LPM to active) & 0.016\,mW \\
\hline\end{tabular}
\label{settings}
\end{table}
\begin{figure}[]
\centering
\includegraphics[width=\linewidth, height=1.2in]{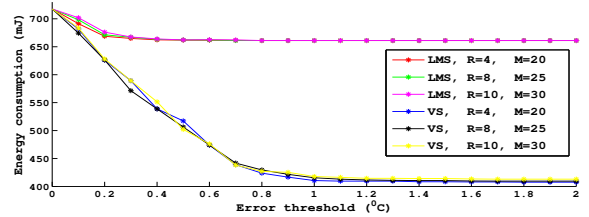}
\caption{Combined energy consumption for one Type-I and one Type-II virtual sensors v/s LMS-based scheme for two sensors.}
\label{fig:energy}
\end{figure}

The system parameters and their settings are listed in Table~\ref{settings} to calculate the energy consumption. The value of each parameter is calculated using methods described in \cite{amiri2010measurements,prabhakar2012zen}. Using these parameters and the number of times a sensor senses and transmits data, energy consumption per node is calculated. As a Type-I VS (dormant node) is always accompanied by a Type-II VS (active node), a combined energy consumption is calculated and compared with the LMS-based method described in \cite{santini2006adaptive}. As the error threshold increases, VSs consume lesser energy at the PSs. Note that energy consumption calculated here is based on only sensing and data transmissions. Using VSF, a Type-I VS can achieve higher energy conservation due to its dormant PS. Furthermore, with lesser number of transmissions, the relay nodes in WSN have to spend lesser energy in forwarding the data packets towards the sink. As a result, further energy consumption is achieved, which is not accounted in this study.
\begin{figure}
\centering
\includegraphics[width=\linewidth]{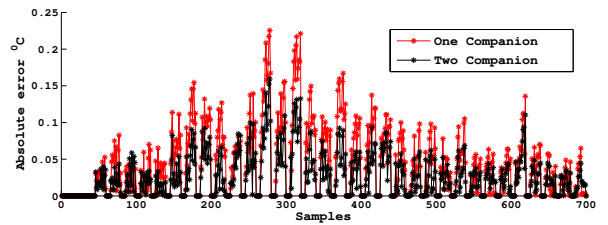}
\caption{Prediction error for a Type-I VS, when one and two active nodes are used as companion.}
\label{fig:multi}
\end{figure}

To improve prediction accuracy of a Type-I VS multiple spatially correlated sensors can be used as companions. In Fig.~\ref{fig:multi}, prediction error based on one and two companions are shown. It is evident that the prediction accuracy improves when multiple spatially correlated sensors are used for prediction. Nevertheless, more energy is consumed due to multiple active PSs that act as companions. There is a trade-off between data accuracy and energy spent in the nodes.

\begin{figure}[]
\centering
\includegraphics[width=\linewidth]{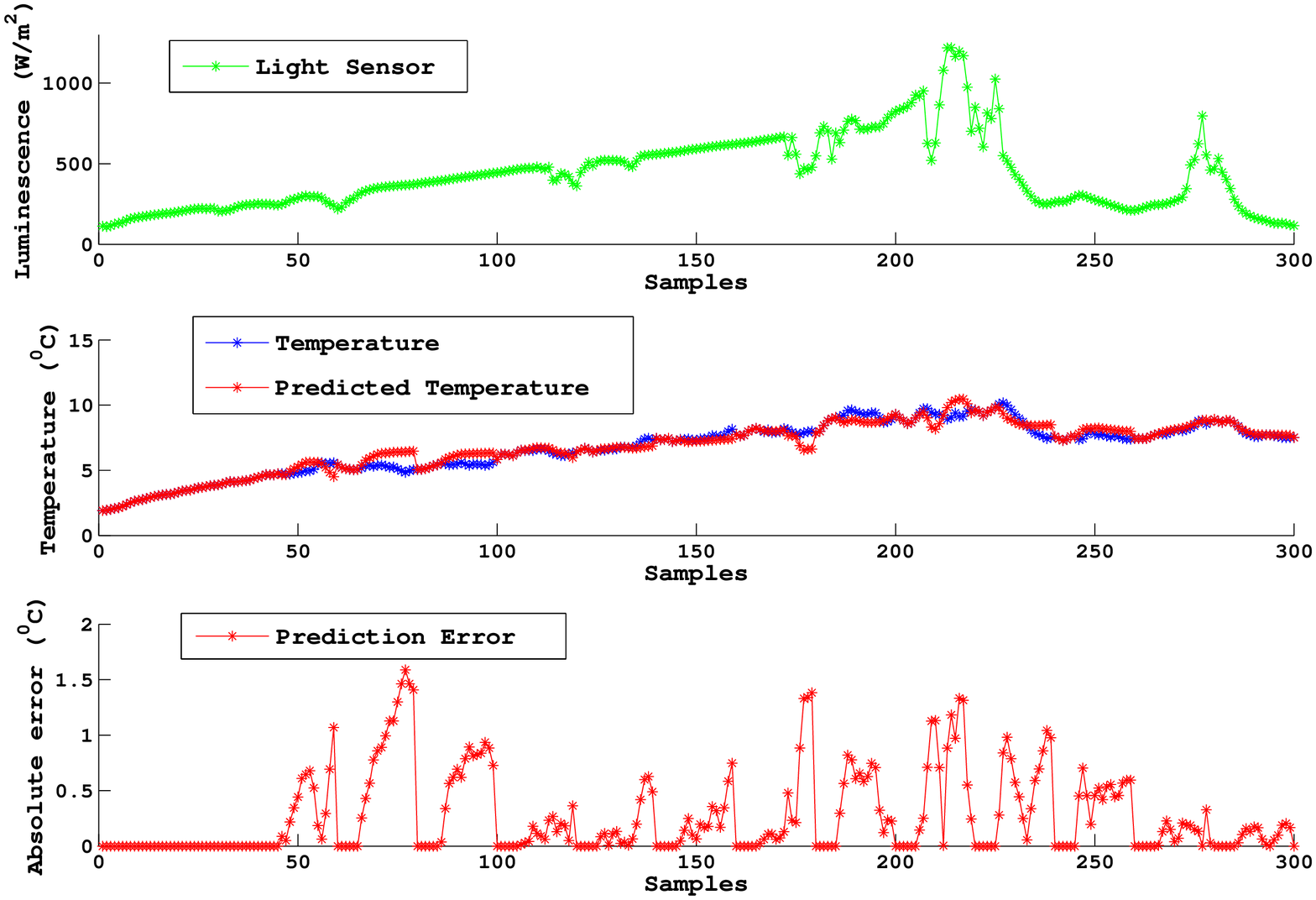}
\caption{Temperature sensor values are predicted using a light sensor as its companion.}
\label{fig:heterogeneous}
\end{figure}

As an extension of VSF, we have also tried \textit{heterogeneous virtual sensing}, where the companion sensor is monitoring a different physical parameter. We have tried to predict temperature using a light sensor (see Fig.~\ref{fig:heterogeneous}). The light and temperature sensor data are collected from~\cite{nrel}. This shows the effectiveness of heterogeneous virtual sensing. However, there are some issues to be addressed such as light sensor data can change quickly over a short period of time, while temperature changes gradually. This affects accuracy of predictions. Fine tuning our technique to adapt to the situation based on estimated prediction errors may help in increasing the reliability and usability of VSF in large scale deployment. Further, VSF could also be used in case some type of sensors is not available at a location. To achieve better error bounds, more investigations are required with respect to heterogeneous virtual sensing. 

\section{Conclusions}
\label{con}
In this article we introduced virtual sensing framework, which can be used in many periodically reporting WSN applications. Moreover VSF can be used in conjunction with intelligent sleep-wakeup schemes to increase energy savings. VSF predicts multiple consecutive sensor data while the physical sensor remains dormant. We have utilized the inherent spatio-temporal correlation amongst the sensor data without having any \textit{a priori} knowledge about the statistics of the data, location of the sensor nodes and type of observed physical parameters. A case in point is predicting temperature with a light sensor with a tolerable error bound. The proposed prediction scheme adapts to the changes in the sensor data. Using our technique, we have achieved a significant improvement on energy saving as compared to other methods while maintaining high accuracy of the sensor data. We have reported around 4\% of error in predicted data.  We believe that our technique will be useful when large number of sensors are deployed and with the advent of Internet of Things (IoT) paradigm. One criticism of this work, as it stands now, is that we have data from both sensors in our VSF. However, the idea is to make sensors sleep. When some sensors sleep, a method to estimate the error between predicted value and what measurement those sensors would have thrown is needed. This is a hard problem but is crucial to achieve higher savings in energy and higher network lifetime.


\section*{Acknowledgment}
This work is supported by an EU FP7 project, called iCore (http://www.iot-icore.eu/), contract number: 287708.



%

\tiny
\bibliographystyle{abbrv}
\bibliography{ncc}
%
%

\end{document}